\documentstyle[prl,aps]{revtex}
\twocolumn
\begin{document}
\draft
\title{Universality in Dynamic Coarsening of a Fractal Cluster}
\author{Baruch Meerson and Pavel V. Sasorov \cite{adr}}
\address{The Racah Institute  of  Physics, The Hebrew
University   of  Jerusalem,
Jerusalem 91904, Israel}
\maketitle
\begin{abstract}
Dynamics of coarsening of a statistically homogeneous 
fractal cluster, created by morphological 
instability
of growth during an early stage of a first order phase transition, is 
investigated theoretically. An exact mathematical setting of the problem, 
obeying a global conservation law, is presented. A statistical
mean field theory is developed that accounts for shadowing 
and assumes that the fractal dimension of the 
cluster is
invariant in time. The coarsening dynamics are found to be self-similar,
and the corresponding dynamic scaling 
exponents are calculated for any Euclidean dimension. 
\end{abstract}
\pacs{PACS numbers: 64.60.-i, 61.43.Hv, 64.60.Ak, 05.70.Fh}
\vskip1pc
\narrowtext

A non-equilibrium dissipative 
system undergoes relaxation to equilibrium after
the driving agent is ``switched off" or depleted. 
In complex systems the relaxation dynamics 
can  be quite complicated, and it is natural to seek for
the dynamic scaling and universality.  An 
instructive, exactly solvable
example of dynamic scaling in
relaxation (coarsening) of rough surfaces with 
nonconservative
dynamics is given by the deterministic
(undriven) KPZ-equation \cite{KPZ,BS}. A much older example
is
decay of homogeneous and isotropic hydrodynamic 
turbulence \cite{Bat,Bar}. Finally, there is an important class of relaxation 
problems 
related to 
phase ordering dynamics, non-conserved and conserved,
in the bulk \cite{Gunton,Voorhees,Bray} and on the surface \cite{Zinke}. 

In the case
of conserved dynamics, ``switching off"
of the driving agent occurs naturally as a result 
of a global conservation law. There is an important 
class of non-equilibrium systems that exhibit strong morphological 
instabilities and ramified growth at an early stage of the dynamics,
show phase ordering at an intermediate stage,
and finally approach a simple thermodynamic 
equilibrium. Examples are provided by realistic diffusion-controlled
growth systems, such as deposition of solute from a supersaturated solution
and solidification from an overcooled
liquid. The stage
of morphological instability and its implications has been under extensive
investigation \cite{Langer,Kessler,Brener,Mineev,BMT}. On the contrary,
the (usually much longer) phase-ordering stage has not been addressed. 
We claim in this Letter that this generic phase-ordering 
stage exhibits intrinsic
self-similarity, calculate the 
scaling exponents for any Euclidean dimension $d>1$ and make 
predictions 
that can be tested in 
experiment or numerical simulations.

We will concentrate, for concreteness, on the problem of growth of
a single nucleus (we will call it cluster) from a supersaturated 
solution and 
start with a 
minimalistic mathematical 
setting that describes correctly the whole dynamics, from the stage of growth
to the final equilibrium.
Let $u ({\bf r},t)$ be the mass concentration 
of the solution normalized to
the (constant) density of solute in the compact solid phase. 
The field $u$
is governed 
by the diffusion equation 
\begin{equation}
\frac{\partial u}{\partial t} = D \nabla^2 u
\label{a}
\end{equation}
in a finite $d$-dimensional domain.
We specify a no-flux boundary condition,
\begin{equation}
\nabla_n u \mid _{\Gamma} = 0
\label{b}
\end{equation}
on the external boundary $\Gamma$, where index $n$ stands for the normal 
component of a vector. Assuming that the moving interphase 
boundary $\gamma$ is in local
thermodynamic equilibrium, we employ the Gibbs-Thomson relation
\begin{equation}
u\mid _{\gamma} = u_0 (1+\lambda_0 \kappa)\,, 
\label{c}
\end{equation}
where  $u_0$ is the (normalized) equilibrium
concentration of the solution in the bulk, $\lambda_0$ is the 
capillary length and $\kappa$ 
is the local curvature
for $d=2$, or the mean curvature for $d>2$. (For simplicity,
we will limit ourselves to an isotropic surface tension.) 
Finally, mass conservation
at the moving boundary yields the well-known relation for the normal speed: 
\begin{equation}
v_n = \frac{D \nabla_n u}{1-u}\mid_{\gamma}\,.
\label{1}
\end{equation}

It is easy to check that Eqs. (\ref{a})-(\ref{1}) preserve the 
total mass of the solute. In the normalized form  
\begin{equation}
\Omega_c + \int_{\Omega} u \, d {\bf r}  = \mbox{const} \,,
\label{2}
\end{equation}
where $\Omega_c$ is the volume (area) of the
cluster, while $\Omega$ denotes the region unoccupied by the cluster. This 
simple conservation law 
describes the depletion of solution during the 
deposition
process and imposes an important constraint on the dynamics. This constraint
does not appear in the more traditional formulations of the
diffusion-controlled growth 
problem \cite{Langer,Kessler,Brener,Mineev,BMT}, where an infinite system
is studied, and the boundary condition 
corresponding to
a constant (positive) flux or constant supersaturation at
${\bf r}\rightarrow \infty$ is used instead of 
Eq. (\ref{b}). Notice that, even in the limit of strong diffusion,
it is
the 
full diffusion equation (rather than its Laplace's equation limit) and no-flux
condition on $\Gamma$ that provide the conservation law. Also, 
the usually small term $u$ in the denominator of 
Eq. (\ref{1}) should be kept to get Eq. (\ref{2}) right. 

Despite a simple formulation, the moving boundary problem 
(\ref{a})-(\ref{1}) is formidable.
In the remainder of this Letter we will present a mean field theory
of coarsening that exploits a strong resemblance of this problem
to the classical problem of Ostwald ripening \cite{Gunton,Voorhees,Bray}, 
accounts for
shadowing and is based on the assumption that
the fractal dimension of the coarsening cluster remains constant on an
interval of scales shrinking with time
\cite{Toyoki,Sempere}.

In a sufficiently large and ``noisy" system (\ref{a})-(\ref{1})
with small initial supersaturation, a fundamental morphological 
instability \cite{Mullins}
leads
to ramification of the moving interface, so that the interface
becomes fractal (DLA-like) \cite{BMT}. Because of mass conservation, the
supersaturation decreases with time. Correspondingly, the surface-tension
effects mitigate and finally switch off the morphological 
instability. The coarsening stage is characterized by the
(properly scaled down) supersaturation
becoming comparable to the mean curvature of the interface. The total
cluster mass asymptotically approaches a constant value $M_0$, and the main 
effect 
at this stage is the diffusion-controlled mass transfer between different 
branches of the cluster that, in view of Eq. (\ref{c}), promotes
growth of larger branches (that is, those with a 
smaller curvature) at the expense
of smaller ones (those with a larger curvature). This
dynamics strongly resembles Ostwald ripening (OR) \cite{Gunton,Voorhees,Bray}. 
The classical problem of OR deals with statistics of many separate ``drops" 
of the new phase, 
and the corresponding statistical mean-field theory \cite{LS}
assumes a negligibly small volume (or area) fraction of the drops, and 
completely
ignores spatial correlations. It is very important that
the dynamic scaling exponent predicted by
the mean-field 
theory \cite{LS} agrees very well with experiments and simulations 
performed under 
conditions of a finite, or even {\it large}  volume
(area) fraction \cite{Gunton,Voorhees,Zinke} and significant
spatial correlations (even when description in terms of individual
drops becomes irrelevant \cite{Bray}). Inspired by this success, we propose
a mean-field theory of coarsening for a statistically homogeneous 
fractal cluster (that is a cluster whose 
fractal dimension is constant in space). 

Assuming for simplicity,
that the
cluster is (statistically) azimuthally (for $d=2$), or
spherically (for $d=3$) symmetric, we introduce 
a time-dependent distribution function $f(R,r,t)$ for the number of 
branches with size $R$
situated at the radius $r$ from the cluster center. 
Looking for self-similarity, we assume a single scaling 
in $R$ (which implies, in particular, that
the typical
branch size scales like its curvature radius) and
a single scaling in $r$ for the whole cluster. An additional 
variable 
is an effective (coarse-grained)
supersaturation field $\Delta u (r,t)=u(r,t)-u_0$. The $r$-dependence 
in the functions
$f$ and 
$\Delta u$
accounts for shadowing and, as we will see later, for 
shrinkage of the coarsening cluster. This dependence
has no analog in the ``conventional" mean-field theory of OR \cite{LS}. 

Neglecting direct coalescence of branches, we write down a
continuity equation
\begin{equation}
\frac{\partial f}{\partial t} + \frac{\partial}{\partial R} 
\left( V_R f \right) = 0\,,
\label{3}
\end{equation}
while for $V_R$ we adopt the well-known relation \cite{LS,2d}
\begin{equation}
V_R=\frac{1}{R} \left(\Delta u - \frac{d-1}{R}\right)\,,
\label{4}
\end{equation}
following from the solution of an idealized problem of the Laplacian growth 
of an isolated {\it spherical}
particle with the boundary condition (\ref{c}) on its interface
(all dimensional coefficients are scaled down). Actually, the precise form
of Eq. (\ref{4}) is unessential for our purposes. What is really important
(and used in the following) 
is that the characteristic normal speed of the boundary (that is, 
expansion/contraction rate of an individual branch) scales like $R^{-2}$, while
the effective supersaturation scales like the characteristic mean 
curvature (see below).

Introduce the (time-dependent) 
mass content of the cluster within the radius $r$:
\begin{equation}
M(r,t) = \frac{d \pi^d}{[\Gamma (\frac{d}{2}+1)]^2}
\int_0^r dr\,r^{d-1} \int_0^{\infty} dR\,R^d \, f(R,r,t)\,,
\label{5}
\end{equation}
where $\Gamma (x)$ is the gamma-function. Mass conservation of 
the coarsening cluster implies that
$M(r_c,t) =M_0=$const, where $r_c $ is the (time-dependent) cluster 
radius.

We are looking for self-similarity:
\begin{equation}
f (R,r,t) = t^{-\mu} \Phi (R t^{-\nu}, r t^{\beta})\,,
\,\,
\Delta u (r,t) = t^{-\alpha} U(r t^{\beta}).
\label{5a}
\end{equation}
Correspondingly, the cluster radius will scale like $r_c(t)=\eta_0 t^{-\beta}$,
where $\eta_0$ is a constant depending on $d$. Notice, that the anzats 
for $f$ assumes 
self-similarity with respect 
to 
each of the {\it two} arguments, $R$ and $r$. Upon substitution,
Eqs. (\ref{3}) and (\ref{4}) yield 
\begin{equation}
\alpha = \nu =1/3
\label{5b}
\end{equation}
independently of the Euclidean
dimension $d$ and
of (still undetermined) $\beta$ and $\mu$. Employing the mass conservation, 
we obtain a 
relation between $\beta$ and $\mu$: 
\begin{equation}
d \beta + \mu - \frac{d+1}{3} = 0\,.
\label{20}
\end{equation}
Now we will use the assumption of invariance of the fractal 
dimension $d_f$
[on an interval of scales $[l_{min}(t), l_{max}(t)]$ that is shrinking 
with time]. It is convenient to work with the mass density of 
the cluster:
\begin{eqnarray}
\rho(r,t) =   \frac{\Gamma(\frac{d}{2}+1)}{d \pi^{d/2} r^{d-1}} 
\frac{\partial M (r,t)}{\partial r}= \nonumber \\
\frac{\pi^{d/2}t^{\frac{d+1}{3} - \mu}}{\Gamma(\frac{d}{2}+1)}
\, \int_0^{\infty}d \xi \,\xi^d \Phi (\xi,\eta)\,,
\label{6}
\end{eqnarray}
where $\eta = r t^{\beta}$. 

For a (statistically 
self-similar) ``mass fractal" with
the fractal dimension $d_f$ one has
$M(r,t) = A(t) r^{d_f}$
which implies a density
\begin{equation}
\rho (r,t) \sim A (t)\, r^{d_f-d} \sim A (t)\, t^{\beta (d-d_f)} \eta^{d_f-d}\,.
\label{7}
\end{equation}
[Since the 
{\it coefficients} of the scaling laws  
(and even 
the shape functions
$\Phi$ and $U$) will remain undetermined, we have omitted 
here and in the following all $d$- 
and $d_f$-dependent
coefficients.] A direct comparison of Eqs. (\ref{6}) and (\ref{7}) yields
\begin{eqnarray}
A (t) \sim t^{\frac{d+1}{3}-\mu + \beta (d_f-d)} \quad
\mbox{and} \nonumber \\
\int_0^{\infty} d\xi \xi^d \Phi(\xi,\eta) \sim \eta^{d_f-d} \,.
\label{7a}
\end{eqnarray}
At scales smaller than $l_{min} (t)$ the cluster is already compact.
It is natural to identify $l_{min} (t)$ with a typical (for example, mean)
value of the branch size $\bar{R}(t)$  that, according to Eq. (\ref{5b}), 
scales with time 
like $t^{1/3}$ \cite{min}.  Therefore, at radii $r<\bar{R}(t)$ 
the mass density must be constant. On 
the other
hand, Eqs. (\ref{7}) and (\ref{7a}) predict 
the following behavior of the mass density at $r=\bar{R}(t)$: 
$$\rho (\bar{R},t) \sim 
t^{\frac{d+1}{3} - \mu} (\bar{R}t^{\beta})^{d_f-d} \sim
t^{-(d-d_f) \beta - \mu + \frac{d_f+1}{3}}\,.$$
Requiring this to be constant, we have
\begin{equation}
(d-d_f) \beta + \mu - \frac{d_f+1}{3} = 0\,.
\label{8}
\end{equation}
Combining Eqs. (\ref{20}) and (\ref{8}) we find 
\begin{equation}
\beta=\frac{d-d_f}{3 d_f}\,, \quad 
\mu=\frac{2 d + 1}{3} -\frac{d^2}{3 d_f}\,.
\label{8a}
\end{equation}
One can see that $\beta>0$ for any $d$, so the coarsening cluster is indeed
shrinking
with time \cite{shrink}. 
Also, $\mu>0$ for any $d$. Having found the dynamic scaling 
exponents $\alpha, \beta, \mu$ and $\nu$,
we can make a number of predictions. A ``global" prediction concerns scaling 
of the 
cluster perimeter (for $d=2$) or interface 
area (for $d=3$). These quantities decrease
like
$t^{-1/3}$ (independently of $d$ and $d_f$!). Another ``global"
prediction concerns the already mentioned 
scaling of the cluster radius: $r_c \sim t^{-(d-d_f)/3 d_f}$. For example, for
a 
DLA-like cluster ($d_f\simeq 1.71$ for $d=2$, and $d_f\simeq 2.5$
for $d=3$), our predictions are $r_c \sim t^{-0.057}$ and $t^{-0.067}$, 
respectively. Returning to 
Eq. (\ref{6}), we see that the mass
density $\rho (r,t) \sim t^\sigma r^{d_f-d}$, where $\sigma =(d-d_f)/3$.
For $d=2$ and $3$ we get $\sigma \simeq 0.10$ and $0.17$, respectively. 

Self-similar phase-ordering 
dynamics of a fractal cluster that we have considered
crosses over to a compact relaxation dynamics on a time 
scale $t_*$ for which $l_{min} (t_*) \sim l_{max} (t_*)$, that
is $\bar{R} (t_*) \sim r_c (t_*) $. Finally, the cluster acquires 
spherical
(circular) shape, corresponding to the only possible two-phase 
equilibrium
in the system \cite{sphere}.

Our mean-field theory is based on the assumption of a statistically
homogeneous cluster.
It is well
known that the diffusion-controlled fractal growth crosses over to a compact 
growth when the 
growing cluster
reaches the`` diffusion length" \cite{BMT}. Our estimates show that,
for a sufficiently large system and small initial supersaturation, the depletion
effects become important before the cluster
radius has a chance to reach the diffusion length. As the result, the cluster 
stops growing
and the coarsening effects that we have described starts dominating. 

A number of refinements of the theory are straightforward (like an 
account for growth
anisotropy by introducing an additional,
angle-dependent factor in the mean-field variables $f$ and $\Delta u$). What
is probably more important is that similar mean-field theories can be developed
for other transport mechanisms (for example, interface-controlled coarsening,
see, {\it e.g.} \cite{Zinke}). The main difference will be in the value
of exponent $\nu$, and scaling arguments yield $\nu=1/2$ in this
limit. In addition, we must put $U(\eta)=const$, as the supersaturation
field is uniform in this case \cite{MS96}.
Once $\nu$ is found, the rest of exponents can be easily calculated. For 
the {\it general}
$\nu$, one obtains
$$ \beta=\frac{\nu(d-d_f)}{d_f}\,, \quad
\mu=\nu \left(2d+1-\frac{d^2}{d_f}\right)\,.$$
Correspondingly,
the interface area (perimeter in 2d) 
will decrease like $t^{-\nu}$ independently 
of $d$ and $d_f$.

Predictions of the mean-field theory
can be checked
in experiment and numerical simulations. 
Simulations 
are possible with different types of models. For the diffusion-controlled
coarsening,
one option is the sharp-interface model (\ref{a})-(\ref{1}). Another
can employ a conserved phase-field model 
\cite{Bray,Hohenberg}.
As noise is not important at the coarsening stage, one can use the noiseless
version of any of these two models and 
start directly from a fractal cluster \cite{interface}. 
One more alternative would be to employ a modified DLA algorithm 
with the
surface tension \cite{Kadanoff} and total 
mass conservation properly taken into account. The simulations can 
be time-consuming,
as the (pronounced) coarsening stage is necessarily much longer than the 
growth stage. 

We are grateful to G.I. Barenblatt for a sharp question and to A. Peleg for
a useful remark. This work was supported in part 
by a grant from Israel Science Foundation, administered 
by the Israel Academy of Sciences and Humanities, and by the Russian Foundation
for Basic Research (grant No. 96-01-01876).



\begin{references}
\bibitem{2d} For $d=2$, the logarithmic divergence inherent in the Laplace  
equation should be taken care of, which does not affect the scaling
exponents, similar to \cite{M}.
\bibitem{min} In doing so, we actually make a natural assumption that
the function $U (\eta)$ is everywhere finite (neither zero, nor
infinite).
\bibitem{shrink} Such a shrinkage was
first predicted in Ref. \cite{Sempere} in the context of 
sintering of fractal matter.
\end{references}
\end{document}